\documentclass [prl, superscriptaddress, showkeys, twocolumn]{revtex4}
\usepackage {graphicx}
\usepackage {bm}
\usepackage {amsmath}
\usepackage {hyperref}
\def\figw{0.5}
\def\LSMO{$\mathrm{La_{0.7}Sr_{0.3}MnO_3}$ }

\begin{document}

\title{
Correction of Non-Linearity Effects
in Detectors for Electron Spectroscopy}

\begin{abstract}
Using photoemission intensities and a detection system employed by
many groups in the electron spectroscopy community as an example, we
have quantitatively characterized and corrected detector non-linearity
effects over the full dynamic range of the system.  Non-linearity
effects are found to be important whenever measuring relative peak
intensities accurately is important, even in the low-countrate regime.
This includes, for example, performing quantitative analyses for
surface contaminants or sample bulk stoichiometries, where the peak
intensities involved can differ by one or two orders of magnitude, and
thus could occupy a significant portion of the detector dynamic range.
Two successful procedures for correcting non-linearity effects are
presented.  The first one yields directly the detector efficiency by
measuring a flat-background reference intensity as a function of
incident x-ray flux, while the second one determines the detector
response from a least-squares analysis of broad-scan survey spectra at
different incident x-ray fluxes. Although we have used one
spectrometer and detection system as an example, these methodologies
should be useful for many other cases.
\end{abstract}

\author{N.~Mannella} 
\altaffiliation{Present address: Physics Department, Stanford University, Stanford, CA, USA.}
\email[Correspondence should be addressed to N. Mannella: ]{nmannella@lbl.gov}
\affiliation{Department of Physics, UC Davis, Davis, CA, USA}
\affiliation{Materials Sciences Division, Lawrence Berkeley National Laboratory, Berkeley, CA, USA}

\author{S.~Marchesini}
\altaffiliation{Present address: LLNL, Livermore, CA, USA}
\affiliation{Materials Sciences Division, Lawrence Berkeley National Laboratory, Berkeley, CA, USA}

\author{A.~W.~Kay}
\altaffiliation{Present address: Intel Corporation, Portland, OR, USA}
\affiliation{Department of Physics, UC Davis, Davis, CA, USA}
\affiliation{Materials Sciences Division, Lawrence Berkeley National Laboratory, Berkeley, CA, USA}

\author{A. Nambu}
\affiliation{Materials Sciences Division, Lawrence Berkeley National Laboratory, Berkeley, CA, USA}
\affiliation{Department of Physics, University of Tokyo, Tokyo, Japan}
\author{T. Gresch}
\affiliation{Materials Sciences Division, Lawrence Berkeley National Laboratory, Berkeley, CA, USA}
\affiliation{Institute of Physics, University of Zurich, Zurich, Switzerland}
\author{ S.-H. Yang} 
\altaffiliation{Present address: IBM Almaden Research Laboratory, San Jose, CA, USA}
\affiliation{Materials Sciences Division, Lawrence Berkeley National Laboratory, Berkeley, CA, USA}
\author{ B. S. Mun} 
\altaffiliation{Present address: Advanced Light Source, LBNL, Berkeley, CA, USA}
\affiliation{Department of Physics, UC Davis, Davis, CA, USA}
\affiliation{Materials Sciences Division, Lawrence Berkeley National Laboratory, Berkeley, CA, USA}
\author{J. M. Bussat} 
\affiliation{Engineering Division, Lawrence Berkeley National Laboratory, Berkeley, CA, USA}
\author{A. Rosenhahn} 
\altaffiliation{Present address: Heidelberg University, Heidelberg, Germany}
\affiliation{Materials Sciences Division, Lawrence Berkeley National Laboratory, Berkeley, CA, USA}

\author{C. S. Fadley}
\affiliation{Department of Physics, UC Davis, Davis, CA, USA}
\affiliation{Materials Sciences Division, Lawrence Berkeley National Laboratory, Berkeley, CA, USA}
\maketitle

\section{Introduction}

Electron detection systems are an integral part of any experimental setup 
for electron spectroscopy. Any deviation from an ideal linear response in 
which the true electron flux incident on the detector is not proportional to 
the response signal of the detector may cause undesirable effects in the 
recorded spectra. Seah and co-workers have previously discussed methods for 
detecting non-linearity effects in photoelectron spectroscopy counting 
systems for spectra measured with laboratory x-ray sources 
\cite{Seah:1992,Seah:1995,Seah:2003}. In this work, we 
develop methods for correcting for such non-linearities in a fully 
quantitative way.

Non-linearity is an ever-present concern in electron spectroscopy 
measurements. With laboratory x-ray excitation sources and solid samples, 
the differences between the highest and lowest photoelectron peak 
intensities can differ by as much as two orders of magnitude. Beyond this, 
for any electron spectrum, measurement of features in the higher-intensity 
low-energy secondary electron tail region of any spectrum can push many 
detection systems into non-linear behavior. For the particular case of 
synchrotron radiation experiments on solids, intensity levels can even more 
easily be found to exceed the linear response range of the detection 
systems. For example, several groups have observed non-linearity effects 
when using state-of-the-art photoelectron spectrometers such as for example 
the Gammadata/Scienta series of spectrometers 
\cite{Kay:2000,Kay:2001,Kay:2002,Garnier:1999,Nilsson:1,Kikas:2000,Finazzi:1}. 
In this situation, non-linearity effects are likely to be present when 
high-cross-section peaks are excited with bright sources (e.g. undulators), 
or even more so in resonant photoemission experiments during which photon 
energy is scanned \cite{Garnier:1999}. For example, prior work on 
multi-atom resonant photoemission (MARPE) by several groups was strongly 
affected by non-linearity effects which produced irregularities in the size 
and shape of the measured resonances, with this effect arising through 
changes in the inelastic background underneath the peak whose intensity was 
being measured 
\cite{Kay:2000,Kay:2001,Kay:2002,Garnier:1999,Nilsson:1,Kikas:2000,Finazzi:1}.

More generally, the possible occurrence of non-linearity effects should 
always be kept in mind whenever measuring relative intensities accurately is 
important, since it is not limited to resonance experiments. In fact, we 
have found for our example system that non-linearity effects are present 
even when the exciting energy is far away from any resonance and the 
countrates are relatively low, of the order of a few KHz. Examples of 
measurements significantly altered by non-linearity effects occurring at low 
countrates include quantitative analysis of complex oxides via core-level 
intensities \cite{Mannella:2003}, relative intensities in angle-resolved 
valence spectra \cite{Chang:1} and dichroism measurements on 
ferromagnetic systems \cite{Wernet:1}.

In this paper, we explore in detail these non-linearity effects using 
photoemission intensities as an example, and focusing in particular on the 
response of the detector over the low-countrate region. We demonstrate two 
quantitatively accurate correction procedures to correct for non-linearity 
effects. The first one directly yields the detector efficiency by measuring 
a flat-background reference intensity as a function of a linearly-varying 
incident x-ray flux, while the second determines the detector response from 
a least-squares analysis of broad-scan survey spectra, each of which spans a 
considerable fraction of the dynamic range, obtained at different incident 
x-ray fluxes. Although we have used one spectrometer system as an example, the Gammadata-Scienta SES200, the methodologies presented should be 
applicable to a broad array of situations. 

\section{Experimental}
\subsection{The Detector System}

\begin{figure}[htbp]
\centerline{\includegraphics[width=\figw\textwidth]{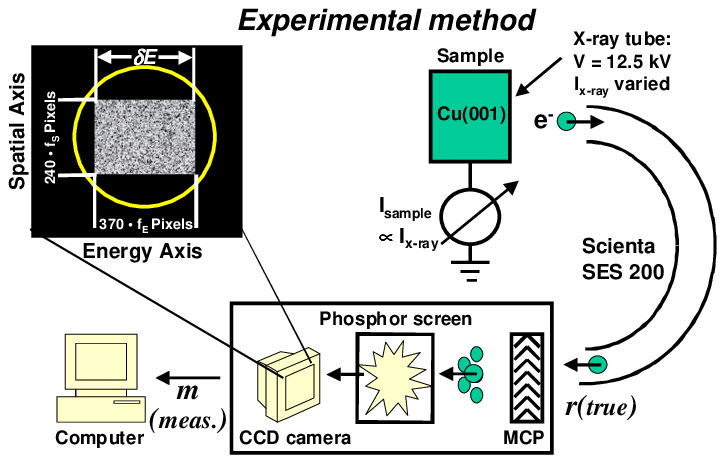}}
\caption{
Schematic illustration of the experimental geometry. The maximum 
active region of the detector (shown on the CCD camera monitor) includes 370 
pixels along the energy axis and 240 pixels along the spatial axis, which is 
reduced to about 70,000 pixels via a rectangular window circumstribed by the 
circular microchannel plates and mating phosphor. Both the energy and 
spatial axes can be gated to include only a specific rectangular portion of 
the detector. The filling fractions $f_{E} $ and $f_{S} $ along the energy 
and spatial axis respectively can be set via software. Note that a linear 
variation in the x-ray emission current (at constant high voltage between 
filament and anode) results indeed in a linear increase in the flux of 
photons at the sample and thus of the electrons incident on the front of the 
MCP. In fact, the sample-to-ground current (in turn proportional to the 
photon flux at the sample), as measured with a picoammeter and recorded as a 
function of the x-ray emission current, has been found to track linearly 
with the emission current of the x-ray source at constant high voltage. 
\label{fig:1}
}
\end{figure}

We have performed our experiments using a Gammadata-Scienta SES200 
spectrometer and detector system, as located on the Advanced Photoelectron 
Spectrometer-Diffractometer situated at the Berkeley Advanced Light Source 
\cite{rtensson:1994,Fadley:1995}. The detection system used is 
that provided by the manufacturer as part of the standard equipment, and is 
schematically illustrated in Fig. \ref{fig:1}. 

A microchannel plate multiplier (MCP) is followed by a phosphor screen
at high voltage in ultra-high vacuum (UHV), so as to convert charge
pulses into visible light pulses. A standard CCD camera
\cite{Manifactured:1}, mounted outside the UHV chamber and focused on
the phosphor screen through a glass viewport, is finally responsible
for recording the light pulses on the phosphor and therefore
performing the actual event detection. We have operated the detector
primarily in the \textit{greyscale} or
\textit{analogue} (GS) mode in which 
integrated CCD charge is used for counting, rather than in the
\textit{black-and-white} or \textit{digital mode} (BW) 
in which individual pulses are counted. 
However, we also present some results based on the BS mode. In the GS
mode, the readout involves a measurement of the collected charge in
the pixel with an 8-bit analog to digital converter (ADC). For the GS
mode, the equivalent of the BW mode discriminator is a digital mask
that eliminates low-order ADC bits in an attempt to discard spurious
noise counts. Further details have been reported elsewhere
\cite{Kay:2000,Kay:2001,Kay:2002}. With 
any change in the discriminator levels or voltages across the MCP, the 
conditions under which the detector measures a count are altered and the 
response function will be modified. Unless otherwise explicitly specified, 
the detector has been intentionally used as delivered and installed by the 
manufacturer, leaving its settings at their recommended value at setup. 
However, in what follows, we will explore the influences of changing some of 
these settings on non-linearity.

This detector is intrinsically two-dimensional. The nature of the
hemispherical energy analyzer to which the detector is attached
results in one pixel axis of the camera representing the electron
kinetic energy. The perpendicular axis, for our purposes, simply
represents multiplexed detection at each energy. These axes will be
referred to as the \textit{energy} and
\textit{spatial} axis respectively (cf. Fig. \ref{fig:1}). 
The camera views the 40 mm diameter circular phosphor screen, with the
rectangle circumscribing this maximum active region of the detector
including about 370 pixels along the energy axis (a number we will
refer to as $N_{E})$ and 240 pixels along the spatial axis (a number
we will refer to as $N_{S})$. Within a square circumscribing the
circle, a maximum fraction, $\pi /4$, of the pixels within the square
will actually include the phosphor screen image, leading to a maximum
of approximately $370 \times 240 \times \pi /4 \approx 70,000$ pixels
available for counting in two dimensions when the camera views the
largest fully-filled rectangular image.

The detector operates in a mode for which both the energy and spatial
axes can be gated to include only a specific rectangular portion of
the detector in the final binned data: we will refer to the fractional
coverage along the energy azis as $f_{E}$ and that along the spatial
axis as $f_{S}$. However, once these limits are selected, all counts
for a given energy axis coordinate (i.e. a line of pixels along the
spatial axis) are summed in hardware and only this binned sum is
available for readout. This sum of spatial-axis pixels for a fixed
energy pixel coordinate is referred to as a
\textit{detector channel}, 
whereas a pixel will refer to one pixel of the CCD camera.

In order to provide a detailed description of the detector response,
the detected signal must first be processed from a typical
distribution of total measured counts to a distribution of measured
countrates\textit{ per pixel}. Once the detector signal has been
acquired, the average countrate per pixel is computed as a function of
the true countrate per pixel, revealing the response of the detector
for the current detector settings (GS--or perhaps BW--mode,
discriminator/mask setting, MCP and phosphor voltages).

The fact that the full camera image cannot be stored for analysis prevents 
the most accurate corrections of the effects to be considered here. That is, 
only in the limit of using a single pixel per detector channel can the 
actual per-pixel countrate be obtained. However, we have dealt with this 
problem by gating the detector so as to have it count over only much smaller 
selected regions, as will be discussed further below.

The detector and analyzer can be run in two different modes, a
\textit{fixed} or
\textit{snapshot} mode as well as a \textit{dithered} 
or \textit{swept} mode. In the fixed mode, the analyzer settings
determine the linear kinetic energy distribution over the energy-axis
of the detector and are held constant. For a given setting $f_{E}$,
the detector will see a kinetic energy range of $\delta E$
(cf. Fig. \ref{fig:1}) that is a maximum of about 10{\%} of the mean
kinetic energy passed by the spectrometer. In this mode, the
per-channel counts, which are actually sums of spatial pixel counts,
are simply stored directly as read from the detector. By using only a
narrow portion of the spatial axis over which the count-rate is nearly
constant, the recorded counts may be trivially converted to a reliable
per-pixel countrate. For this particular case of data collected in the
fixed mode, the correction from \textit {per-channel} counts $M$ to
countrate per pixel $m$ is given by
\begin{equation}
\label{eq:1a}
\tag{1a}
m = \frac{M}{\tau \cdot f_{S} \cdot N_{S} }\,,
\end{equation}
where $\tau $ is the total dwell or counting time of the
spectrum. Or, if we illuminate the detector with a uniform flux of
electrons, then $m$ can be obtained from the total countrate over all
channels $T$ via:
\begin{equation}
\label{eq:1b}
\tag{1b}
m = \frac{T}{\tau \cdot f_{E} \cdot N_{E} \cdot f_{S} \cdot N_{S} }\,,
\end{equation}\setcounter{equation}{1}
where $\tau$ is again the total dwell or counting time of the
spectrum, $N_{E} = 370$ is the maximum number of active pixels along
the energy-axis, $f_{E} $ is the fraction of the detector that is
filled along the energy axis, $N_{S} = 240$ is the maximum number of
active pixels along the spatial axis, and $f_{S}$ is the fraction of
the detector that is filled along the spatial axis
(cf. Fig. \ref{fig:1}). Here, we have assumed that the filled portion
of the spatial axis has essentially uniform countrate over the summed
pixels. If this uniformity condition is not met, the efficiency may
vary significantly across the spatial axis of the detector, and 
Eq. (\ref{eq:1a}, \ref{eq:1b})
will give only some sort of average spanning a part of the dynamic
range of the detector. The requirement of having uniform illumination
over the active area of the detector can be experimentally achieved by
using only a narrow portion of the detector along both the spatial and
energy axes. The region of the detector over which counts are
accumulated, indicated via the percentage of each of the two axes over
which counting is permitted ($f_{E}$ and $f_{S}$, along the energy and
spatial axis, respectively), can be adjusted via software. A previous
investigation provided evidence that there is no significant change in
the response function over any evenly illuminated surface of the
detector, permitting us to much simplify the procedure for correcting
spectra \cite{Kay:2001,Kay:2002}. 
Normally, photoemission experiments are performed in a
\textit{dithered} or \textit{swept} mode that involves sweeping the
kinetic energy of the electrons accepted by the analyzer so that all
energies in the final spectrum are accumulated in sequence by each
channel in the detector. This is primarily done to allow parallel
detection channels to be used while eliminating the channel-to-channel
differences in the detector gain in the final spectra.  For the
dithered mode the correction from measured per-channel counts $M$ to
an average countrate per pixel $m$ is given by an equation similar to
\eqref{eq:1a}: 
\begin{equation}
\label{eq:2}
m = \frac{M}{ \tau ' \cdot f_{S} \cdot N_{S} }\,,
\end{equation}
where $\tau'$is the total time that each pixel has spent in counting at 
each energy channel, as summed over the total no. of sweeps \cite{The:1}.

\subsection{Experimental methodology for detector characterization}
In order to determine the response of the detector, one needs to determine 
the measured countrate as a function of the true countrate. To accomplish 
this in the most direct way, the true countrate, which is proportional to 
the number of electrons incident on the front face of the MCP, must be 
adjusted in a controlled manner while recording the measured countrate at 
the detector. Once the detector signal has been acquired, the average 
measured countrate per pixel is computed as a function of the true countrate 
per pixel, revealing the response of the detector for the used detector 
settings (GS or BW mode, discriminator/mask setting, MCP and phosphor 
voltages).

In this study, we have used electrons emitted during the photoemission
process (photoelectrons) as a source of true
countrates. Photoelectrons were provided by exciting with a standard
laboratory x-ray source a Cu (110) single crystal in an \textit{as-is}
uncleaned condition, i.e. containing a stable amount of contamination
in the UHV environment of the experiment. It is only important that
the sample is in a stable condition during the duration of the
measurements. We have in the present study used a standard x-ray tube
(un-monochromatized dual-anode Al K$_\alpha $/Mg K$_\alpha $, Perkin
Elmer Model 04{\-}548), which has a power supply permitting variable
emission power which can be adjusted in 1 $W$ steps at fixed high
voltage.

We note that there is a fundamental question as to whether a linear
variation in the x-ray emission current (at constant high voltage
between filament and anode) results indeed in a linear increase in the
flux of photons at the sample and thus of the electrons incident on
the front of the MCP. We thus verified initially that the total
electron current from the sample tracked linearly with the emission
current of the x-ray source at constant high voltage. The
sample-to-ground current (in turn proportional to the photon flux at
the sample) was measured with a picoammeter and recorded as a function
of the x-ray emission current, and thus also power since the voltage
has been held constant (cf. Fig. \ref{fig:1}). This relationship is
found to be quite linear over the range of x-ray power used in this
study (5-300 W), with all quadratic or higher-order terms contributing
less than 5{\%} of the linear component within this range, as already
shown in a previous investigation \cite{Seah:1995}. Thus, using either
the sample current or the x-ray power as a measure of the true
countrate introduces negligible differences in the final response
function analysis.
\section{Results And Discussion}
\subsection{The detection of non-linearity effects}
Ideally, the behavior of the detector as a function of the true
countrate should be completely linear. In this case, the detector
response would be described as $m(r) = \varepsilon \cdot r$, where
$m(r)$ and $r$ denote the measured and the true countrate per pixel
respectively, and $\varepsilon $ is a counting efficiency factor. The
constant $\varepsilon $ would thus be equal to one in an ideal system,
but it is for us only necessary to know it to within some constant
factor. When the detector deviates from the ideal behavior, the
detector response must be described by
\begin{equation}
\label{eq:3}
m(r) = \varepsilon (r) \cdot r\,,
\end{equation}
where $\varepsilon (r)$ can be termed the \textit{efficiciency
function} or \textit{detector response function}, and it now depends
on the true countrate, reflecting the deviation from ideal
behavior. In order to correct measured countrates into true countrates
it is necessary to determine the response function of the detector and
invert Eq. \eqref{eq:3}.  We note that since the signal is detected
after being processed by the CCD camera, the values for the measured
countrate (and, consequently, also for the true countrate) are not
absolute, but are determined by the particular choice of the detector
parameters (for example, discriminator threshold settings).  Before
discussing the procedures for the quantitative determination of this
response function, we comment on a couple of straightforward ways to
\emph{detect} non-linearity effects by making use of survey spectra 
measured at different x-ray fluxes.

\begin{figure}[htbp]
\centerline{\includegraphics[width=\figw\textwidth]{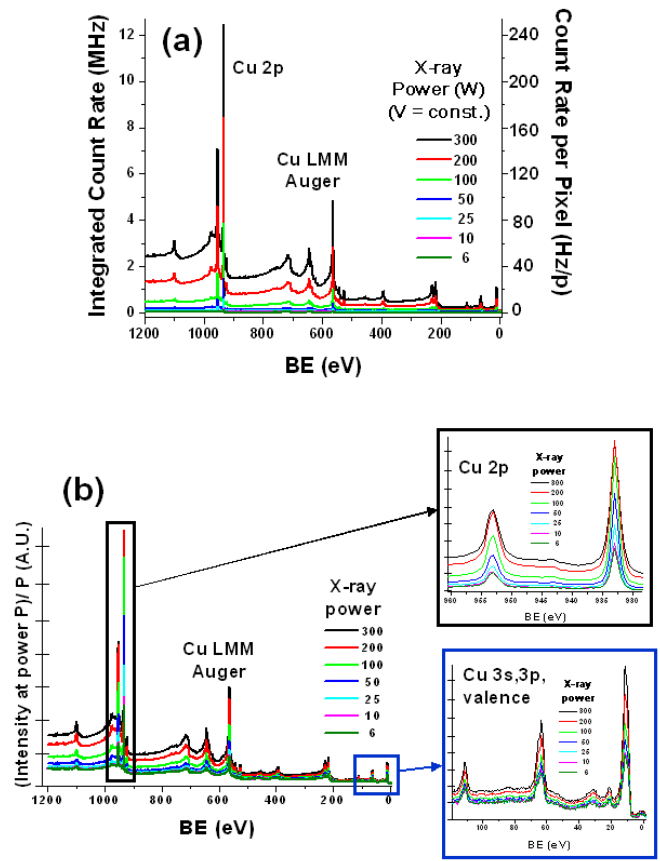}}
\caption{
(a) Broad-range survey spectra collected in the dithered (swept) 
mode from a Cu (110) sample, as excited by Al K$\alpha $ ($h\nu $ = 1486.6 
eV) radiation. Some more intense spectral features are labeled. Some weaker 
peaks result from a Ta clip at the edges of the Cu sample. The left ordinate 
here is an integrated rate assuming that 50,000 pixels count at the rate per 
pixel given on the right scale, and is this only appropriate to a situation 
of uniform illumination of the detector. (b) The same spectra as shown in 
(a) after they have been normalized with respect to x-ray flux. The fact 
that the spectra do not lie on top of one another provides unambiguous 
evidence for the presence of non-linearity effects.
\label{fig:2}
}
\end{figure}

In Fig. \ref{fig:2}a we show broad-range survey spectra collected in
the dithered mode from a Cu (110) sample, as excited by Al K$\alpha $
($h\nu = 1486.6$ eV) radiation.

These spectra span countrates ranging from a few KHz to $ \approx  12$
MHz, corresponding to countrates per pixel in the range $m \approx  1
- 240$ Hz for the detector active area we have used. For our
conditions of gating the active portion of the detector via $f_{S}$
and $f_{E}$, the total no. of active pixels is thus about 50,000, a
number we will use in estimating total maximum uniform countrates
later \cite{The:2}. The same spectra are shown in Fig. \ref{fig:2}b
after they have been normalized with respect to the x-ray fluxes. If
the detector were linear, all spectra in Fig. \ref{fig:2}b should lie
on top of one another, but it is evident that they are not, with
factors of up to 4 separating them in the higher intensity regions at
higher binding energy, as illustrated more quantitatively in the upper
inset showing the Cu 2p spectral region. Even within the narrow
binding energy range of 0 to 120 eV (lower inset), there can be
differences of a factor of 2-3.

Another direct way to monitor non-linearity effects in electron
detector systems makes use of \textit{ratio plots}, as introduced by
Seah and co-workers \cite{Seah:2003}, which consist of ratios of
intensities in survey spectra measured at different x-ray fluxes. In
Fig. \ref{fig:3}, we plot the ratios of the uncorrected intensities of
the individual spectral points collected at values of the x-ray power
set to 300, 200, 100, 50 and 25 W and the intensity of the same
spectral points in energy collected at 25 W. All intensities have been
normalized by dividing by their respective x-ray emission currents,
and the ratios are plotted versus the intensities of the relevant
numerator spectrum. We note that the ratio plots are completely
equivalent to plotting the ratios of the efficiencies curves as a
function of the measured countrate. In fact, if two spectral points
are recorded at two different emission currents related by a scaling
factor $n$, from Eq. \eqref{eq:3} we can write the ratio of the efficiencies as:
\begin{equation}
\label{eq:4}
\frac{\varepsilon (nr)}{\varepsilon (r)} = \frac{m(nr)}{n \cdot m(r)} = 
{\left( {\frac{m(nr)}{nr}} \right)} \mathord{\left/ {\vphantom {{\left( 
{\frac{m(nr)}{nr}} \right)} {\left( {\frac{m(r)}{r}} \right)}}} \right. 
\kern-\nulldelimiterspace} {\left( {\frac{m(r)}{r}} \right)}\,,
\end{equation}
which shows that this ratio is equal to the ratio of the intensities
of the two spectra normalized by the respective x-ray emission
currents at which they have been collected. Ideally, the efficiency
would have a constant value, so that the ratio in Eq. \eqref{eq:4} should be
constant. The main effect observed in Fig. \ref{fig:3} is that the
ratios of the efficiency of the detector increase for measured
countrates per pixel up to 60 - 70 Hz and decrease for countrates per
pixel greater than 90 Hz, while in an ideal system one would expect
these curves to be horizontal straight lines lying on top of one
another, most simply of value unity, as shown in the figure.

We now consider two different methods for determining the response
function of the detector and correcting non-linearity effects. The
first method directly yields the response function by measuring a
flat-background reference intensity as a function of incident x-ray
fluxes, while the second method determines the response function from
what is effectively a least-squared-fit analysis of broad-scan survey
spectra taken at different incident x-ray fluxes.

\subsection{Correction method 1: measurement of flat-background 
reference intensity as a function of incident x-ray flux}

The most obvious way to determine the response function of the
detector is to record directly the measured countrate at the detector
while adjusting the true countrate in a controlled manner. This is
easily accomplished by measuring a flat-background region in a
spectrum from a sample with a stable surface while varying the
incident x-ray flux.

\begin{figure}[htbp]
\centerline{\includegraphics[width=\figw\textwidth]{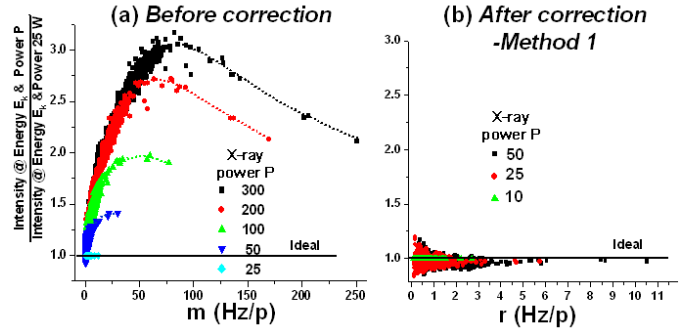}}
\caption{
 Ratio plots (cf. ref. \cite{Seah:2003}) of spectral intensities at a
 given kinetic energy and for different x-ray powers (equivalent to
 fluxes). The solid line indicates the behavior of an ideal linear
 detector with unit efficiency. (a) is before correction, and (b)
 after correction by Method 1. All points have been referred to the
 lowest power of 25 W (a) and 10 W (b). After correction, the ratio
 plots look like horizontal straight lines lying on top of one
 another, as expected in an ideal system. The deviation from a
 horizontal straight line shown for countrates approaching zero is
 simply due to the higher fractional statistical uncertainty that is
 typical of a Poisson distribution.
\label{fig:3}
}
\end{figure}

Although photoemission experiments are usually performed in a dithered
mode, this mode was not used here because the inherent averaging over
the detector would be detrimental to the analysis of the detector
behavior. The analyzer and detector were on the contrary run in a
fixed mode. Here we stress the importance of having uniform
illumination over the active area of the detector in order not to have
significant variations of the detector efficiency across the spatial
axis of the detector. This requirement was experimentally achieved by
setting via software the region of the detector over which counts were
accumulated equal to 20{\%} and 40{\%} of the spatial and energy axes
respectively. The use of a featureless region of the spectrum (for
example, for the Cu (110) sample shown in Fig. \ref{fig:2}a, suitable
regions would correspond to the binding energy (BE) ranges 134.6-164.6
eV and 850-900 eV) along with the use of a gated (40{\%}) portion of
the energy axis allows one to be able to measure several detector
channels at the same time providing better statistics. The countrate
per pixel can then be derived from Eq. \eqref{eq:1a}. For some of our
measurements, a similar flat region in the spectrum from a \LSMO
  sample was used in order to achieve
higher intensities, as discussed in more detail below.

Once the detector signal has been acquired and converted to countrate
per pixel, this method yields directly the response function of the
detector as a function of the x-ray emission current or power, which
is in turn proportional to the true countrate, as discussed before. By
changing the operational mode of the analyzer (e.g. pass energy and
slit size) of the analyzer, it was possible to derive the detector
response in different regions of its dynamic range, thus permitting
the measurement of various portions of the response function of the
detector, particularly the one corresponding to less than 5 Hz per
pixel. As previously shown \cite{Kay:2001,Kay:2002}, the 
only effect introduced by changing the settings of the analyzer is
simply a multiplication of the true countrate by a constant scaling
factor. The nature of the scaling factor is immaterial to this
discussion, but it must be compensated for in order to properly and
self-consistently determine per-pixel countrates.

\begin{figure}[htbp]
\centerline{\includegraphics[width=\figw\textwidth]{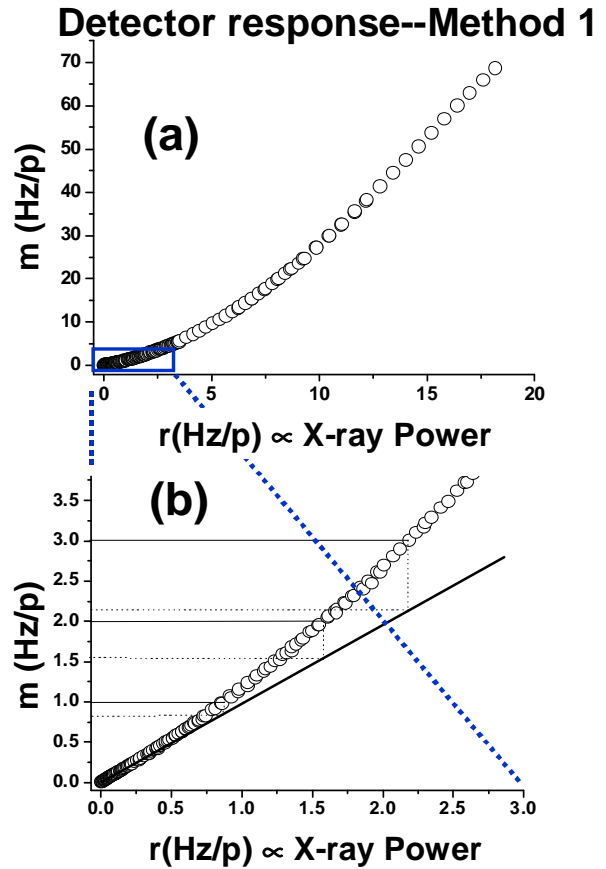}}
\caption{
(a) The detector response measured via the correction procedure of
Method 1. Note the ``quadratic'' deviation from linearity at low
countrates.  (b) The detector response function for very low measured
countrates (less than 3 Hz per pixel). Note that non-linearity effects
are already present at measured countrates as low as 1 Hz per pixel.
\label{fig:4}
}
\end{figure}

Within each setting of the detector operational mode (e.g. GS or BW)
and other detector and analyzer settings, the x-ray power was varied
in the range 5-300 W (at fixed constant voltage V = 12.5 kV). We first
combined several measurements of different portions of the response
function corresponding to different operational mode settings into an
overall measurement of the GS mode response function up to a measured
rate of about 70 Hz per pixel (corresponding to a maximum total
countrate over all energy channels of 3.5 MHz), as shown in
Fig. \ref{fig:4}. The data shown in Fig.
\ref{fig:4} have been 
taken with a \LSMO single crystal (containing
a stable amount of contamination in the UHV environment of the
experiment) and photoelectrons emitted from a featureless region with
BE range 440 -- 480 eV. This was done to obtain a measured rate of 70
Hz per pixel, e.g. higher than the 35 -- 40 Hz per pixel which could
be obtained from the Cu (110) sample, (cf. Fig. \ref{fig:2}a).

Note that we generally do not know the point (if any) at which the
measured and true countrates exactly coincide. In Fig. \ref{fig:4}, we
have arbitrarily set the true countrate scale so that the measured and
true countrates are the same for count rates approaching zero, with
the asymptotic behavior of the measured detector response as the true
countrate goes to zero being a straight line with slope equal to
unity. This choice is of course equivalent to set the efficiency equal
to unity at zero true countrate. It is important to realize that this
choice is arbitrary, and that it does not affect the results of any
correction we make.

Fig. \ref{fig:4} shows that the detector responds with
 significant deviations from the 
ideal linear behavior described by Eq. \eqref{eq:3}, 
\emph {even at very low 
countrates}. This type of non-linearity for low countrates can
approximately be described as a quadratic deviation from linearity, as
previously observed for this particular detector in both GS and BW
modes
\cite{Kay:2002}. In particular, an inspection 
of Fig. \ref{fig:4}b shows that the detector starts already to deviate
from an ideal behavior at 0.5 Hz per pixel or a maximum countrate of
25 kHz. If we quantify the deviation from linearity as in $d = [m(r) -
r] / r\times 100$, these data show that $d$ = 19, 29 and 40 {\%} for
measured countrates equal to 1, 2 and 3 Hz per pixel, respectively.

Once the detector response function is determined, only a simple
interpolation algorithm is needed to invert Eq. \eqref{eq:3} so as to
express the true countrates as a function of the measured
countrates. It should also be noted that only after a spectrum has
been corrected from measured to true countrate is a photon-flux
normalization appropriate.

\begin{figure}[htbp]
\centerline{\includegraphics[width=\figw\textwidth]{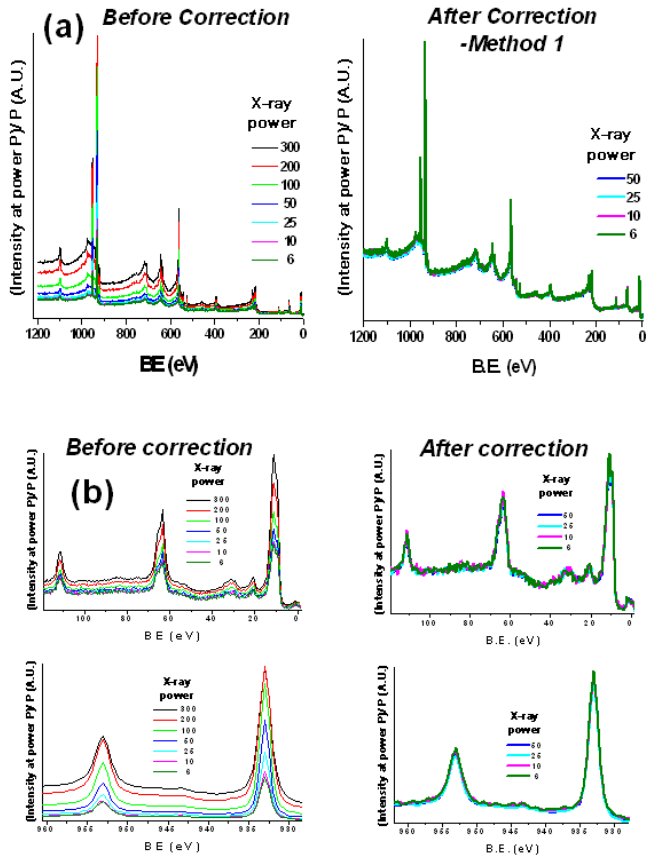}}
\caption{
The same spectra shown in Fig \ref{fig:2}b, but compared before and after 
detector non-linearity correction via Method 1. In (a), the full spectra are 
shown, and in (b) certain blowup regions. Note the limited power range 
possible with Method 1, going only up to about 70 Hz/pixel as measured.
\label{fig:5}
}
\end{figure}

Finally, we show in Fig. \ref{fig:5} the same spectral comparisons as
in Fig.
\ref{fig:2}b, 
but with and without the correction applied: it is clear that all
normalized spectra for different fluxes coincide to a high accuracy
(within 4.5 {\%} for all data points) after correction. Also,
Fig. \ref{fig:3}b makes the same point via the ratio plots. In fact,
after correction, the ratio plots look like horizontal straight lines
lying on top of one another, as expected in an ideal system. The
deviation from a horizontal straight line shown for countrates
approaching zero is simply due to the higher fractional statistical
uncertainty that is typical of a Poisson distribution, which scales as
the inverse of the square root of the counts. These results thus
provide unambiguous evidence that the above-described procedure yields
the correct determination of the response function and is effective in
correcting non-linearity effects.

Nevertheless, an inspection of Figs. \ref{fig:3}b and \ref{fig:5}
reveals a minor inconvenience of this method. It has been possible to
correct over the whole binding energy range only the spectra taken
with x-ray powers of less than 50 W at most, corresponding to a
maximum countrate of $ \approx 70$ Hz per pixel or $
\approx  3.5$ MHz maximum total countrate over all pixels. The cause for 
this limited range lies in the impossibility of finding a suitable
featureless region in the spectra whose countrate is high enough to be
able to drive the detector over a wider dynamic range
\cite{Changing:1}.  Moreover, the necessity of adjusting in a
controlled manner the incident x-ray flux can in principle impose
stringent requirements on existing experimental setups. For example,
standard x-ray tubes are not necessarily equipped with a power supply
which allows quasi-continuous variation of the emission x-ray power
(at constant voltage) with a stepsize as small as a few
watts. Finally, combining several measurements of different portions
of the response function of the detector by changing different
operational mode settings into an overall measurement of the GS mode
response function can be time-consuming. As an example, the collection
of the several data sets combined in Fig. \ref{fig:4} took about 24
hours. These considerations have motivated the development of an
alternative procedure for correcting non-linearity effects that we now
describe below.

\subsection{Correction method 2: Analysis of broad-scan survey 
spectra at different incident x-ray fluxes}

The possibility to develop a new correction procedure was initially
triggered by observing that a single broad-scan spectrum can provide
in a single measurement a highly dense set of measured countrates. For
example, Fig. \ref{fig:2}a shows that the survey spectrum taken with
the power set to 300 W yielded in a few minutes a distribution of
measured countrate ranging from 0 to 240 Hz per pixel. The possibility
of determining the detector response function from an analysis of
survey spectra is thus appealing since it permits sampling a wide
portion of the detector response in a relatively short amount of time.

To make this idea more quantitative, consider a set of $N$ survey
spectra measured on the same sample, but with different incident
fluxes $n= n_{1}, n_{2}, \dots ,n_{N}$ as shown in Fig. \ref{fig:2}a.
For the case of data collected in the dithered mode, the one used to
acquire the broad-scan survey spectra, the correction from per-channel
count-rate $M$ to count-rate per pixel $m$ is given by
Eq. \eqref{eq:2}. When expressed in countrate per pixel, the survey
spectra provide a distribution of measured countrates $m = m(n_{j}
,E_{k} )$ for a given x-ray flux $n_{j}$ and kinetic energy $E_{k}$ of
the photoelectrons.

The true countrates per pixel $r(n_{j} ,E_{k} )$ for a given x-ray
flux and kinetic energy of the photoelectrons can now be expressed as
a polynomial expansion of order $P$ of the measured countrates per
pixel $m(n_{j},E_{k} )$ with real coefficients $a_{i}$:
\begin{equation}
\label{eq:5}
r(n_{j} ,E_{k} ) = \sum\limits_{i = 1}^{P} {a_{i}  m^i(n_{j} 
,E_{k} )} ,\,
\begin{cases}
 \forall\, j = 1,2,\dots N   \\
 \forall\, k = 1,2,\dots Q ,
\end{cases}
\end{equation}
where $Q$ denotes the number of equally-spaced kinetic energy values used to 
collect the spectra. Here, we have set the coefficient $a_0$ (which represents the dark
current background in the absence of any excitation) equal to zero, as
this background is often negligible or can simply be measured and
subtracted from all the measurements, but one can simply extend the
summation to the 0 order term if necessary to include this. The
determination of the response function and the correction of
non-linearity effects are thus reduced to the computation of the
unknown coefficients $a_{i}$.

The \textit {normalized} true countrates do not depend on the incident 
flux, such that we can write:
\begin{equation}
\label{eq:6}
\frac{r(n_{1} ,E_{k} )}{n_{1} } = \frac{r(n_{j} ,E_{k} )}{n_{j} },\,
\begin{cases}
 \forall\, j = 2,3,\dots N   \\
 \forall\, k = 1,2,\dots Q .
\end{cases}
\end{equation}

From Eqs. \eqref{eq:5} and \eqref{eq:6}, we can thus write out a system of
$N - $1 equations as:
\begin{equation}
 \frac{1}{n_{1}}\sum\limits_{i = 1}^{P} {a_{i} 
\cdot m^i(n_{1} ,E_{k} )} = \frac{1}{n_j} 
\sum\limits_{i = 1}^{P} {a_{i} \cdot m^i(n_{j} ,E_{k} )} \,,\\
\label{eq:7}
\end{equation}

As in our treatment of the first method, we have arbitrarily set the
measured and true countrates to be the same at countrates approaching
zero, which corresponds via this limit to setting the arbitrary value
$a_{1} = 1$.  This still leads to a completely general result for the
response function, since the true and measured counts can differ by an
arbitrary factor. From Eq. \eqref{eq:7} we then have with trivial
rearrangement, another system of equations:
\begin{widetext}
\begin{equation}
 \underbrace {\vphantom{\sum\limits_{i = 2}^{P}}
\frac{m(n_{j} ,E_{k} )}{n_{j}} - 
\frac{m(n_{1} ,E_{k} )}{n_{1} }}_{\bm {B}} 
 = \underbrace {\sum\limits_{i = 2}^{P} 
{\left[ {\frac{m^i(n_{1} ,E_{k} 
)}{n_{1} }-\frac{m^i(n_{j} ,E_{k} )}{n_{j} }} \right]}}_{\bm{C}} \cdot 
\underbrace {\vphantom{\sum\limits_{i = 2}^{P}} a_i}_{\bm{A}},
\begin{cases}
 \forall\, j = 2,3,\dots N   \\
 \forall\, k = 1,2,\dots Q  .
\end{cases}
\label{eq:8}
\end{equation}
\end{widetext}
In the matrix and vector notation introduced above, ${\bm{A}}$ is
a $(P - 1)$-long column vector, $\bm{B}$ is a $Q\times (N -
1)$-long column vector, and ${\bm{C}}$ is a $Q\times (N - 1)$ by
$(P - 1)$ matrix. Thus, in the ideal case for which there is no
statistical error in the experimental data, ${\bm{B}} - {\bm{CA}}= \bm{0}$,
 and it would represent an over-determined set of equations
for determining the coefficients $a_{i}$.

For the actual case with statistical variations in the data,
Eq. \eqref{eq:8} thus describe an over-determined system of $Q\times (N
- 1)$ linear equations in the unknown $(P - 1)$ coefficients $a_{i}$'s
that can be solved for maximum likelihood by minimizing 
$\left|\bm{B} - \bm{CA} \right|^2$, i.e. solving the 
\textit{normal equation }of the linear least-squares problem:
\begin{equation}
\label{eq:9}
\nabla _{{\bm{A}}} \left| {{\bm{B}} - {\bm{CA}}} \right|^2 = 
2{\bm{C}}^T{\bm{CA}} - 2{\bm{C}}^T{\bm{B}} = 0\,,
\end{equation}
where the superscript $T$ denotes the transposition operation. The polynomial 
coefficients embedded in $A$ can be obtained by standard methods such as the 
LU decomposition \cite{Press:1} or simple matrix inversion as:
\begin{equation}
\label{eq:10}
{\bm{A}} = \left( {\bm{C}}^T{\bm{C}} \right)^{ - 1} \cdot 
\bm{C}^T \bm{B}\,,
\end{equation}
\noindent
where $\mathbf{C^{T}C}$ is a small $(P-1)$ by $(P-1)$ matrix. 
For better numerical precision in the matrix inversion, we have 
rescaled the measured countrates per pixel $m$ to vary from 0 to 1. Note that 
this overall approach is analogous to fitting the assumed polynomial form to 
the experimental data via a least-squares criterion. In practically 
implementing this scheme, we find that including powers up to $P \approx  
12$ is necessary. The values of the other parameters for the results shown in 
this work are $N = 7$ and $Q = 2400$. 

\begin{figure}[htbp]
\centerline{\includegraphics[width=\figw\textwidth]{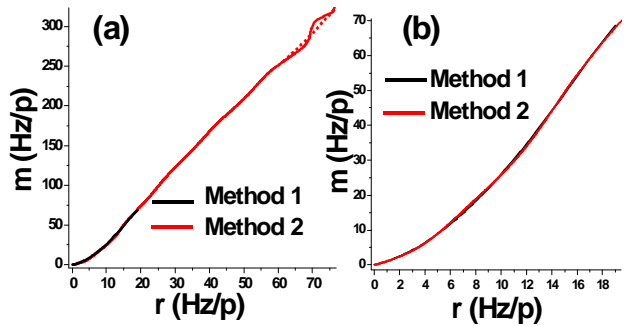}}
\caption{
Comparison of the detector response determined with correction 
Methods 1 and 2. In (a), a broad range going up to above 300 Hz/pixel is 
shown, and in (b) only the more limited region in which the two methods 
overlap.
\label{fig:6}
}
\end{figure}

We have applied this fitting procedure to all the data points
belonging to the survey spectra shown in Fig. \ref{fig:2}a. The
detector response function has in this way been determined for
measured countrates up to over 250 Hz per pixel (Fig. \ref{fig:6}a),
approximately a factor of 4 higher than the range accessed by the
first method. Fig. \ref{fig:6}b also shows that there is excellent
agreement between Methods 1 and 2 over the much narrower range covered
by Method 1. We stress that in order to be able to invert
Eq. \eqref{eq:3}, so as to determine the true counts $r$, the
relationship $m$ vs. $r$ must be a one-to-one mapping.

There is some deviation from a simple smooth curve in
Fig. \ref{fig:6}a above measured rates of approximately 260 Hz per
pixel. This is simply due to the limited number of data points with
countrate per pixel greater than the 250 accessible by our
measurements. However, the smooth dashed curve in this region should
permit correcting even up to about 325 Hz per pixel, corresponding to
a total maximum countrate of 16.25 MHz.

\begin{figure}[htbp]
\centerline{\includegraphics[width=\figw\textwidth]{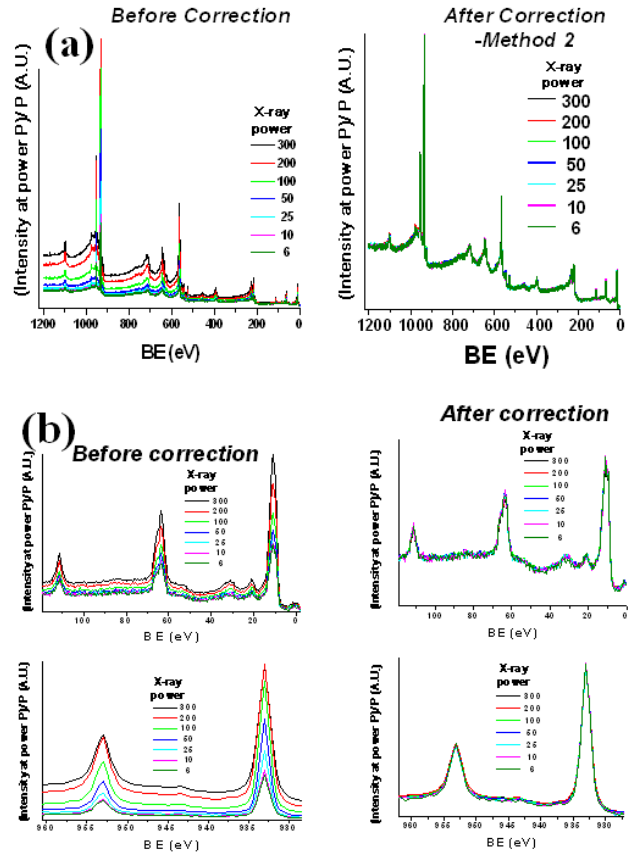}}
\caption{
 As Fig. \ref{fig:5}, but with correction via Method 2--minimum likelihood 
fitting of a polynomial to broad scan spectra via Eqs. \eqref{eq:5}-\eqref{eq:10}. Note the 
similarity with Fig. \ref{fig:5}, indicating that both correction procedures coincide 
and are effective in correcting for non-linearity effects. Note also the 
much broader power range for Method 2.
\label{fig:7}
}
\end{figure}

We show in Fig. \ref{fig:7} the same spectra as in Fig. \ref{fig:2}b,
but again comparing spectra with and without the correction procedure
applied, this time via Method 2. All normalized spectra for different
fluxes coincide to a high degree of accuracy (within at most 6 {\%}
for all data points) over the entire range of the measured countrates
accessed by the spectra shown in Fig. \ref{fig:2}a.  This second
correction procedure is thus very effective in correcting
non-linearity effects and yields the correct determination of the
response function for measured countrates extending to 250 Hz (or even
300 Hz) per pixel, approximately a factor of 4 higher than the range
accessed by the first procedure.

\begin{figure}[htbp]
\centerline{\includegraphics[width=\figw\textwidth]{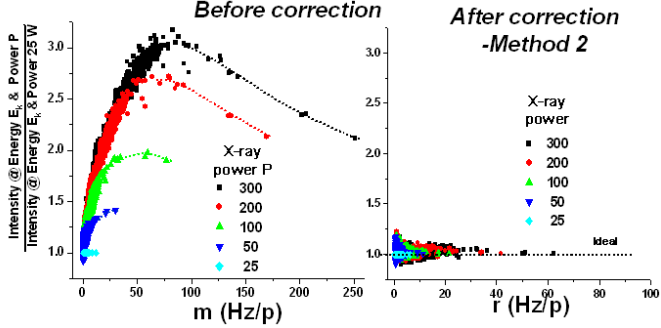}}
\caption{
As Fig \ref{fig:3}, but with spectra compared (a) before and (b) after 
detector non-linearity correction via Method 2. 
\label{fig:8}
}
\end{figure}

Finally, we show in Fig. \ref{fig:8} the ratio plots shown in 
Fig. \ref{fig:3}, but with the 
second correction procedure applied. After correction, the ratio plots look 
like one would expect in an ideal system, with the curves being horizontal 
straight lines lying on top of one another.

We suggest that this procedure will be particularly useful for existing 
experimental set-ups, such as those with standard x-ray tubes equipped with 
a power supply which can allow only a few x-ray emission current settings.

\subsection{Further considerations}

The correction procedures applied above clearly demonstrate successful and 
consistent methods for dealing with a non-ideal behavior in the response 
function of the detector. Maintaining a uniform illumination over the active 
portion of the detector screen is an essential condition for the 
effectiveness of both of the correction procedures described above. It 
should also be noted that there are alternate versions of the 
Gammadata-Scienta hardware that do allow full two-dimensional images to be 
retained in both energy and space and read out from the electronics 
interface. With these systems, it should be possible to apply the correction 
procedure developed here with even greater precision than that demonstrated 
here.

We have also successfully applied both correction procedures described above 
with variable photon flux provided by synchrotron radiation. In this 
particular situation, the variation in the photon flux at the sample can be 
monitored by recording the natural decay of the ring current (although this 
does not normally allow for more than a factor of 3 or so change in flux), 
or by changing the entrance (or exit) slits of the beamline while measuring 
either the photon flux along the beamline with a conventional 
``I$_{0}$''mesh or more directly the sample-to-ground total-electron-yield 
current. However, we point out that caution should be exercised to insure 
that changing the slits does not change the ratio between the beam spot size 
at the sample and the actual sample area seen by the 
spectrometer, otherwise a linear variation of the photon flux at the sample 
may not result in a linear increase of the number of electrons incident on 
the front of the MCP.

\begin{figure}[htbp]
\centerline{\includegraphics[width=\figw\textwidth]{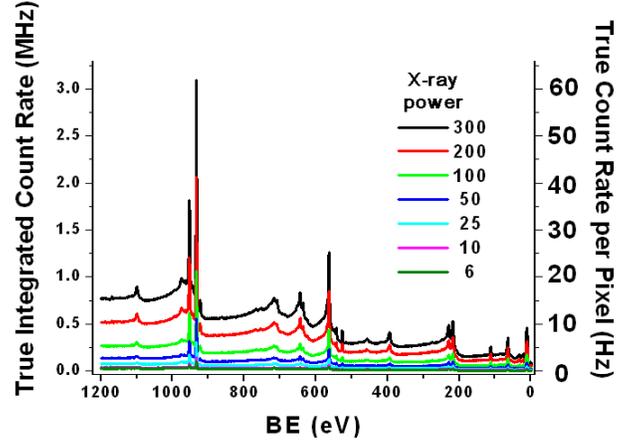}}
\caption{
The same spectra as in Fig. \ref{fig:2}(a), but after they have been corrected 
via Method 2. Note in particular that the maximum countrates per pixel or as 
integrated over all pixels after correction are a factor of 4 less than 
before the correction. The integrated total countrate assumes 50,000 active 
pixels uniformly illuminated, and may be optimistic in estimating maximum 
countrates achievable with this detector in the sense that spectra often 
have high countrate peaks well above background.
\label{fig:9}
}
\end{figure}

\begin{figure}[htbp]
\centerline{\includegraphics[width=\figw\textwidth]{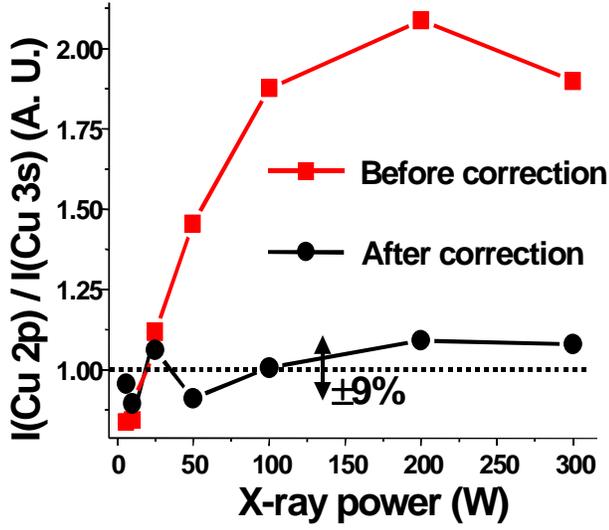}}
\caption{
The intensity ratio of the Cu 2p and Cu 3s core level spectra after 
allowing for the different photoelectric cross sections and electron 
inelastic attenuation lengths, as well as the transmission function of the 
analyzer, so as to yield a number that should in principle equal unity. Note 
the strong flux dependence of the ratio for the case of non-corrected 
spectra, a clear indication of the presence of non linearity effects.
\label{fig:10}
}
\end{figure}

We stress that non-linearity effects should always be kept in mind for any 
case where measuring relative peak intensities accurately is important. As 
one illustration of this, we show in Fig. \ref{fig:9} the same spectra presented in 
Fig. \ref{fig:2}a after the correction procedure has been applied. 
The overall countrate $M$ and the countrate per pixel $m$ now range from 0 to 3 MHz and 60 Hz, 
respectively, that is, a factor of 4 less in range than before the 
correction has been applied. As a more concrete example of how quantitative 
analysis could be affected, Fig. \ref{fig:10} 
shows the ratio of the intensities of 
the Cu 2p and Cu 3s core level spectra, after taking into account 
differences in photoelectric cross section, electron attenuation lengths, 
and the transmission function of the analyzer, so as to effectively be 
taking a ratio of the Cu atomic density via two different spectra from the 
same atom. After the correction, as expected, this ratio is constant and 
equal to 1 within a full range of $\pm $ 9{\%}, while before the correction 
it shows a strong x-ray flux dependence and a value ranging from 1 to a 
little over 2.

As a final point, we note that all of the data reported to this point have 
been obtained with the detector intentionally used as delivered and 
installed by the manufacturer, leaving its settings at their recommended 
values at setup. We also note that several other groups appear to have 
encountered the same type of non-linearity with these standard settings 
\cite{Garnier:1999,Nilsson:1,Kikas:2000}. 
According to the manufacturer's recommendations, a too-low discriminator 
level introduces noise, while a too-high discriminator level influences the 
detection efficiency of low intensity signals and therefore modifies the 
linearity of the intensity scale. The manufacturer recommends setting the 
discriminator by minimizing the dark counts; making sure that the dark 
counts are barely visible is thus thought to ensure that the discriminator 
is not set too high. As we discuss immediately below, we have in fact as 
part of this study varied both the discriminator setting and the phosphor 
and MCP high voltages, but the general type of non-linearity discussed here 
persists.

\begin{figure}[htbp]
\centerline{\includegraphics[width=\figw\textwidth]{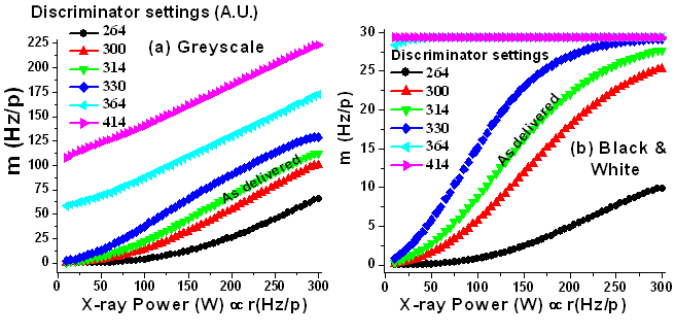}}
\caption{
The effects of changing the discriminator setting on the detector 
response function for the grey scale mode (a) and the black-and-white mode 
(b). All curves here show significant deviations from linearity.
\label{fig:11}
}
\end{figure}

\begin{figure}[htbp]
\centerline{\includegraphics[width=\figw\textwidth]{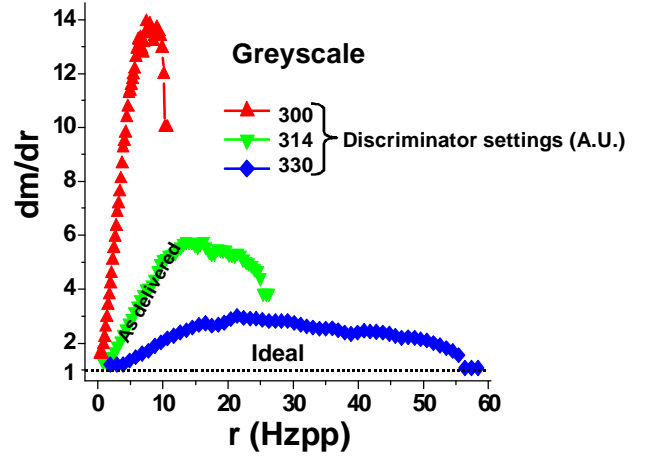}}
\caption{
The derivative of the detector response dm/dr obtained in the 
grey-scale mode (Fig. \ref{fig:11}a) corresponding to the discriminator values of 300, 
314 and 330. If the detector were linear, these should be horizontal lines.
\label{fig:12}
}
\end{figure}

We thus now address the question of whether it is possible that the 
non-linearity effects so far observed are related to poor settings of the 
discriminator level. Seah et al. have in fact pointed out that the 
discriminator setting in a detector very similar to ours can be used to 
improve linearity in certain countrate ranges, although this procedure is 
not expected to eliminate nonlinearity effects over a broad countrate 
range, especially in GS operation due to the nature of this mode 
\cite{Seah:1992,Seah:1995,Seah:2003}. 
In order to investigate whether an improper adjustment of the discriminator 
level on the detector could be held responsible for the non-linearity 
effects here reported, we thus studied the response function for various 
detector discriminator settings in both for the GS and the BW modes. Making 
use of the first method described above, we determined the 
response functions corresponding to six different settings of the 
discriminator, with the results presented in Fig. \ref{fig:11}. 
We deliberately used 
values for the discriminator setting (here reported as numbers in arbitrary 
units) lower and higher than that set by the manufacturer, which was equal 
to 314, so as to investigate what the effect of increasing or decreasing 
the threshold level would be. It is evident from an inspection of 
Fig. \ref{fig:11} 
that there is no value for the settings that we tried that yields the 
correct linear behavior over the entire countrate range accessed by our 
measurements. For the particular case of the GS mode, the discriminator 
settings which would allow one to measure spectra with quasi-linearity are 
those corresponding to the values 300, 314 and 330, centered on the 
manufacturer's recommended setting. Outside of this range, for the value 
equal to 264 we obtained a multi-valued response function, while for values 
equal to 364 and 414 the dark counts are so high that they would constitute 
an unacceptable noise level in the recorded spectra. For values equal to 
300, 314 and 330, all three response functions show quadratic behavior at 
low countrates, as already pointed out before. In order to better quantify 
the deviation from an ideal linear behavior, we show in Fig. \ref{fig:12}
 (for the 
particular case of the GS mode) the detector responses corresponding to the 
discriminator values set to 300, 314 and 330 after they have been 
differentiated, i.e. $dm/dr$. Consistent with our prior analysis, we have 
arbitrarily set the true countrate scale so that the measured and true count 
rates are the same for countrates approaching zero. The non-linearity 
affecting these response functions is evident: in an ideal case, these plots 
should be straight lines with zero slope. Therefore, these results indicate 
that it is generally unlikely that different discriminator settings would 
eliminate non{\-}linearity effects for this particular detection system over 
a broad range of measured countrates normally accessed by typical 
photoemission measurements. Nonetheless, the setting of 330 in this figure 
is somewhat better than the 314 of the standard setup, even though it still 
shows a slope change of about a factor 3 over the range studied.

Beyond this, exploring optimum settings for detector high voltages, phosphor 
high voltages, and discriminators on our detection system is part of routine 
optimization of this detector, an operation performed approximately once a 
year on our system in collaboration with the manufacturer's engineers. We 
have recently verified that, just after re-optimizing (and in fact 
increasing) the MCP voltage, the detector still shows the same type of 
non-linearity, both qualitatively and quantitatively.

The detector response has also proven to be very stable over time. We 
obtained excellent reproducibility of the detector response (and hence 
correction procedure) even after 1 year, thus suggesting that the correction 
procedure does not have to be derived more often than once every six months 
or so, provided that the system is always operated at the same UHV 
conditions and the focus of the CCD camera is not changed. 

Although a precise determination of the true source(s) of the non-linearity 
in this particular detector is beyond the scope of our paper, we briefly 
comment on the possible causes for the non-linearity effects we observe, 
first at high countrates and then at low countrates. Saturation of the 
detector at high countrates occurs most likely because of photon ``pile-up'' 
at the phosphor plate and/or saturation of the CCD camera due to its maximum 
sampling rate. The photon ``pile-up'' at the phosphor occurs when the decay 
time of the phosphor is not sufficiently fast; for the phosphor used in our 
system the decay time is 10 $\mu $s, and thus we can estimate that pileup in 
a given pixel will begin to occur at about ten times the phosphor decay time 
or the equivalent rate of 10$^{4}$ Hz per pixel, a value which is much 
higher than any countrate measured in our work. The saturation of the CCD 
cameral is clearly shown in Fig. \ref{fig:11}b for the spectra taken in 
black-and-white mode with the discriminator set to 364 and 414; in fact, for 
this particular model saturation occurs at 30 Hz per pixel (in 
black-and-white mode) since this is the CCD sampling rate.

It is not clear at the moment what the precise cause of the quadratic 
non-linearity at low countrate is, even though our investigation suggests 
that the most plausible source of these effects is the CCD camera. A too-low 
MCP high voltage would cause the gain to change sensitively as the flux 
change, giving rise to non-linearity effects. Nonetheless, we rule out a 
too-low MCP high voltage as a cause for the quadratic non-linearity, since 
after increasing the MCP the same non-linearity effects are found, as 
explained above. 

It has been suggested that a change in the CCD camera might improve the 
behavior \cite{Johnson:1} of the detector, and this is a direction for 
future investigation. Plausible causes for the non-linear behavior at low 
countrates are CCD dark signal and CCD pattern noise. The first one is 
caused by some leakage currents which would produce charge in some of the 
pixels. It is expected that changing the discriminator values would suppress 
this source of noise, but our measurements (cf. Fig. \ref{fig:11})
 reveal that for 
different discriminator settings non{\-}linearity effects are always 
present. CCD pattern noise refers to any pattern of counts (e.g. hot spots) 
which does not change significantly from frame to frame and, thus, even if 
not properly a random noise, can produce a dark signal-like background, 
however differing from random noise in that it would be dependent on the 
specific location of the CCD pixels used. As already noted above, we do not 
find any evidence on heterogeneity in the behavior of the pixels from one 
part of the detector to another, thus suggesting that CCD patttern noise 
cannot be held responsible for the non-linearity. From our investigation we 
conclude that most likely the cause of the non linearity is the inherent use 
of a CCD camera, since such devices are well known to be non-linear devices 
\cite{Reibel:2003}, with the determination of the precise cause being 
object of future investigation.

More generally, this study constitutes a motivation for improving existing 
detectors and developing new detectors that overcome problems related to 
non-linearity effects over much larger countrate ranges.

\section{Conclusions}

We have developed two procedures for accurately correcting non-linearity 
effects in detectors for electron spectroscopy that should be applicable to 
a broad range of systems. The first one directly yields the detector 
efficiency by measuring a flat-background reference intensity as a function 
of incident x-ray flux, while the second one determines the detector 
response from a least-squares analysis of broad-scan survey spectra at 
different incident x-ray fluxes. To illustrate our correction procedures, we 
have characterized the detector response over a broad dynamic range of a 
state-of-the-art electron spectrometer system (Gammadata/Scienta SES200), 
using photoemission intensities as an example. Although we have studied only 
one spectrometer and detection system, our conclusions and general methods 
for determining and correcting for non-linearity are useful for many other 
cases. For the particular case studied here, our results demonstrate the 
occurrence of ``quadratic'' non-linearity effects which affect the detector 
response function at even very low countrates, far from saturation. Such 
non-linearity effects should thus always be kept in mind for any case where 
measuring relative peak intensities accurately is important, even at low 
countrates. Our results indicate that changing the discriminator settings 
does not eliminate these non-linearity effects, nor does adjusting the 
voltage across the multichannel plates. Finally, this study points out the 
importance of developing new detectors with a linear behavior over the 
entire countrate range accessed by typical experiments in electron 
spectroscopy.

This work was supported by the Director, Office of Science, Office of
Basic Energy Sciences, Materials Science and Engineering Division,
U.S. Department of Energy under Contract No. DE-AC03-76SF00098.


\begin{thebibliography}{17} 
\bibitem{Seah:1992} M.~P.~Seah, M.~Tosa,
Surface and Interface Analysis 18 (1992) 240.
\bibitem{Seah:1995} M.~P.~Seah, Surface and Interface Analysis
23 (1995) 729-732.  
\bibitem{Seah:2003} M.~P.~Seah,
I.~S.~Gilmore, S.~J.~Spencer, J. Electron
Spectrosc. Relat. Phenom. 104 (1999) 73-89. The methods to
check for linearity is described as ISO 21270:2003, ``Surface chemical
analysis - X-ray photoelectron spectrometers and Auger electron
spectrometers - Linearity of intensity scale''.  
\bibitem{Kay:2000}
A.~W.~Kay, Ph.D. dissertation, UC Davis, 2000.  
\bibitem{Kay:2001}
A.~W.~Kay, S.-H.~Yang, E.~Arenholz, B.~S.~Mun, N.~Mannella, Z.~Hussain,
M.~A.~Van Hove and C.~S.~Fadley, J. Electron
Spectrosc. Relat. Phenom.  114, (2001) 1179-1189.
\bibitem{Kay:2002} A.~W.~Kay, F.~J.~G.~de Abajo, S.-H.~Yang, E.~Arenholz,
B.~S.~Mun, N.~Mannella, Z.~Hussain, M.~A.~Van Hove and C.~S.~Fadley, ,
\prb 63, (2001) 115119.  
\bibitem{Garnier:1999}
M.~G.~Garnier, N. Witkowski, R. Denecke, D. Nordlund, A. Nilsson,
M. Nagasono, N. M{\aa}rtensson, and A. F\"{o}hlisch, Maxlab Annual
Report for 1999, Lund, Sweden, and private communication.
\bibitem{Nilsson:1} A. Nilsson, R. Denecke\textit{, et al.}, private
communication.  
\bibitem{Kikas:2000} A. Kikas, E. Nommiste, R. Ruus,
A. Saar, and I. Martinson, Solid State Communications 115,
(2000) 275, and A. Kikas, private communication.  
\bibitem{Finazzi:1}
M. Finazzi and N. Brooke, private communication; G. Paolucci and
K. Prince, private communication.  
\bibitem{Mannella:2003}
N. Mannella, Ph.D. dissertation, UC Davis, 2003.
\bibitem{Chang:1} I. --D Chang and D. Dessau, private communication.
\bibitem{Wernet:1} P. Wernet, N. Mannella, B. S. Mun, S.-H. Yang and
C. S. Fadley, unpublished.  
\bibitem{rtensson:1994} N. M{\aa}rtensson,
P. Baltzer, P.A. Br\"{u}hwiler, J.-O. Forsell, A. Nilsson,
A. Stenborg, and B.Wannberg, Journal of Electron Spectroscopy and
Related Phenomena 70, (1994) 117.  
\bibitem{Fadley:1995}
C.S. Fadley et al., J. Electron Spectrosc. Relat. Phenom. 75,
(1995) 273.  
\bibitem{Manifactured:1} Manifactured by Sony, model
XC-77 E72675, DC 10.5 -- 15 V, 2.2 W.  
\bibitem{The:1} The time $\tau'$ is a little more complicated 
to calculate, but if the active detector window in kinetic energy is
 $\delta E$ cf. Fig. \ref{fig:1}, then the smallest reasonable step in kinetic
energy (or equivalently energy channel width) will be 
$dE =\delta E/(f_{E} N_{E})$. If the spectral region
to be scanned is $\Delta E$ in width, (where $\Delta E$ is usually 
$\ge \delta E$), then, in order to have each detector pixel
contribute equally to each energy channel, the detector has to be
scanned over a range of $\Delta E + \delta E$, involving a
maximum no. of energy steps 
$S = (\Delta E + \delta E)/dE = 
(\Delta E + \delta E) f_{E} N_{E}/\delta E$. If the
actual accumulation time at each energy step is $\delta \tau
$, not including any time necessary for the saving of data and
settling in of the power supplies, and the spectrum is swept
$F$ times, then the total time $\tau'$ to accumulate
a swept-mode spectrum will be $\tau' =FS\delta \tau$.
 If, as is often the case, a certain number of detector energy
channels $\alpha $ is binned together to make a final
spectral energy channel of width $\alpha dE$, where
$\alpha > 1$  and need not be an integer, then the number of
energy steps is reduced to $S' = S/\alpha$ and the total
time becomes $\tau' = FS \delta \tau /\alpha $.
\bibitem{The:2} The filling fractions $f_{E} $ and $f_{S} $ along
the energy and spatial axis have been set to 0.8 and 0.7,
respectively. Consequently, one has $0.8 \cdot 370 \cdot 0.7 \cdot 240
\approx 50000$ as a conversion factor between countrate per pixel and
overall maximum total countrate ($T/\tau $), provided the
detector sees a uniform illumination (see Eq. \eqref{eq:1b}).
\bibitem{Changing:1} Changing the sample could help in some cases,
even though it is unlikely to find a featureless region which can
provide countrates as high as 240 Hz per pixel.  
\bibitem{Press:1}
W.H. Press, S. A. Teukolsky, W.T. Vetterling, B.P. Flannery,
\textit{Numerical Recipies in C: the art of scientific computing} (2nd
ed.), Cambridge University Press, 1992 
\bibitem{Johnson:1}
P. D. Johnson, private communication.  
\bibitem{Reibel:2003}
Y. Reibel, M. Jung, M. Bouhifd, B. Cunin and C. Draman,
Eur. Phys. J. AP 21, 75-80 (2003).  
\end{thebibliography}
\end{document}